\documentclass[aps,prl,reprint,preprintnumbers,showpacs,showkeys,superscriptaddress,floatfix,nofootinbib]{revtex4-1}

\usepackage{graphicx}
\usepackage{epsfig,amssymb,amsmath,color}
\usepackage{epstopdf}
\usepackage{times}
\usepackage{subfigure}

\usepackage[colorlinks=true,citecolor=blue,linkcolor=blue]{hyperref}

\usepackage{ulem}

\begin{document}

\title{Nano-Hz gravitational wave signature from axion dark matter}

\author{Naoya Kitajima}
\email{naoya.kitajima.c2@tohoku.ac.jp}
\affiliation{Frontier Research Institute for Interdisciplinary Sciences, Tohoku University, Sendai, 980-8578 Japan}
\affiliation{Department of Physics, Tohoku University, Sendai, 980-8578 Japan}

\author{Jiro Soda}
\email{jiro@phys.sci.kobe-u.ac.jp}
\affiliation{Department of Physics, Kobe University, Kobe 657-8501, Japan}

\author{Yuko Urakawa}
\email{yuko@physik.uni-bielefeld.de}
\affiliation{Department of Physics and Astrophysics, Nagoya University, Chikusa, Nagoya 464-8602, Japan}
\affiliation{Fakult\"at f\"ur Physik, Universit\"at Bielefeld, 33501 Bielefeld, Germany}

\begin{abstract}
We calculate the accurate spectrum of the stochastic gravitational wave background from U(1) gauge fields produced by  axion dark matter.
The explosive production of gauge fields soon invalidates the applicability of the linear analysis and one needs nonlinear schemes.
We make use of numerical lattice simulations to properly follow the nonlinear dynamics such as backreaction and rescattering which gives important contributions to the emission of gravitational waves.
It turns out that the axion with the decay constant $f \sim 10^{16}$ GeV which gives the correct dark matter abundance predicts the circularly polarized gravitational wave signature detectable by SKA.
We also show that the resulting gravitational wave spectrum has a potential to explain NANOGrav 12.5 year data.
\end{abstract}

\preprint{KOBE-COSMO-20-16, TU-1111}
\maketitle

\paragraph{Introduction}
--
The cosmological observations have uncovered the existence of
 invisible matter known as dark matter.  An axion, originally proposed to solve the strong CP problem~\cite{Peccei:1977hh,Peccei:1977ur,Weinberg:1977ma,Wilczek:1977pj}
 and later explored in a more general paradigm~\cite{Svrcek:2006yi,Arvanitaki:2009fg}, is a 
 candidate of dark matter. To identify dark matter as an axion among various candidates, we need to look for its unique signature. 
 Recently, the NANOGrav 12.5-year data~\cite{Arzoumanian:2020vkk} reported a potential signal of a new physics.
 One may wonder if it is a hint of axion dark matter.
 In this paper, we explore the possibility that axion dark matter can be searched from a measurement of nano-Hertz (nHz) gravitational waves (GWs) through pulsar timing observations such as PPTA~\cite{Hobbs:2013aka}, NANOGrav~\cite{Arzoumanian:2018saf}, and EPTA~\cite{Lentati:2015qwp} . In the next decades, SKA~\cite{Janssen:2014dka} is expected to measure the GWs with an increased precision by several orders of magnitude. 

The axion with a non-zero initial misalignment starts to oscillate, when the Hubble parameter becomes comparable to the mass, behaving as dark matter. Around the commencement of the oscillation, the U(1) gauge fields can be explosively produced through the coupling with the axion  \cite{Garretson:1992vt,Finelli:2000sh}, leading to the emission of sizable amount of stochastic GWs as studied in \cite{Machado:2018nqk,Salehian:2020dsf,Ratzinger:2020koh,Namba:2020kij} \footnote{This mechanism is also applicable to generate primordial magnetic field \cite{Turner:1987bw,Garretson:1992vt,Fujita:2015iga,Adshead:2016iae,Patel:2019isj}, to reduce the abundance of QCD axion \cite{Agrawal:2017eqm,Kitajima:2017peg} and provide the correct relic abundance of the dark photon dark matter \cite{Agrawal:2018vin,Co:2018lka,Bastero-Gil:2018uel}. }.

In this letter, conducing the numerical lattice simulation, which properly solves the nonlinear dynamics of the gauge fields and the axion, 
we show that the axion with the decay constant $f \sim 10^{16}$ GeV can generate the nHz GWs detectable by SKA  while consistently explaining the dark matter abundance.
In addition, it also turns out that the resultant GW spectrum exhibits asymmetry between the two circular polarization modes as a characteristic signature of this scenario.
The reported NANOGrav 12.5 year data can also be explained in this scenario if one requires a drastic suppression for the relic axion density.

\paragraph{Gauge field production from axions}
-- Let us consider the interacting system of an axion ($\phi$) and massless U(1) gauge fields ($A_\mu$).
The Lagrangian density is given by 
\begin{eqnarray}
{\cal L} = -\frac{1}{2}\partial^{\mu} \phi \partial_{\mu} \phi -V(\phi)
  -\frac{1}{4} F^{\mu\nu} F_{\mu\nu} 
  -\frac{\alpha}{4f} \phi F_{\mu\nu} \tilde{F}^{\mu\nu}\!,
\end{eqnarray}
where $F_{\mu\nu} = \partial_\mu A_\nu - \partial_\nu A_\mu$ is the field strength tensor, $\tilde{F}^{\mu\nu} = \epsilon^{\mu\nu\rho\sigma} F_{\rho\sigma}/(2\sqrt{-g})$ is its dual with $g$ being the determinant of the metric, $\alpha$ is a dimensionless coupling constant and $f$ is the decay constant of the axion. 
We consider the potential of the axion, $V(\phi)$, given by
\begin{eqnarray}
V(\phi) = \Lambda^4 \left[  1 -  \cos\left(\frac{\phi}{f}\right)    \right] \ ,
\end{eqnarray}
where $\Lambda$ is a dynamical scale, which relates the axion mass, $m$, to the decay constant as $m= \Lambda^2/f$. 

In what follows, we assume the radiation-dominated flat-FRW Universe as the background spacetime
with the scale factor $a(t) \propto t^{1/2}$. The equation of motion for the axion is given by
\begin{equation} \label{eq:axion_eom}
\ddot{\phi}+3H\dot{\phi}-\frac{1}{a^2} \nabla^2 \phi + \frac{\partial V}{\partial\phi} = -\frac{\alpha}{4f} F_{\mu\nu}\tilde{F}^{\mu\nu},
\end{equation}
with the Hubble parameter $H=\dot{a}/a$. The overdot denotes the time derivative. The variation of the action with respect to $A_0$ yields the constraint equation (modified Gauss's law)
\begin{equation}
\partial_i \dot{A}_i -\frac{\alpha}{fa} \epsilon_{ijk} \partial_i \phi \partial_j A_k = 0,
\end{equation}
and that with respect to $A_i$ leads to the evolution equation of the gauge fields,
\begin{equation} \label{eq:gauge_eom}
\ddot{A}_i+H\dot{A}_i -\frac{1}{a^2} \nabla^2A_i +\frac{1}{a^2} \partial_i \partial_j A_j = \frac{\alpha}{fa} \epsilon_{ijk}(\dot\phi \partial_j A_k - \partial_j \phi \dot{A}_k),
\end{equation}
where we have chosen the temporal gauge, $A_0=0$. 
We assume that 
the axion is homogeneously produced by misalignment mechanism with initial angle $\theta_i = \phi_i/f$ which is set before or during inflation\footnote{
In general, the initial value of the axion fluctuates but such initial fluctuations, as long as they are sufficiently small, do not affect the dynamics of the gauge fields since the self-resonance is inefficient in the cosine potential case~\cite{Fukunaga:2019unq,Fukunaga:2020mvq}.
However, if the axion potential deviates from the cosine type, the self-interaction can induce resonant amplifications of axion fluctuations \cite{Soda:2017dsu,Kitajima:2018zco}, potentially affecting 
%whose growth rate can be larger than that for the gauge fields In that case, 
the dynamics of the gauge fields.}.

As shown below, the interaction with the axion amplifies the gauge fields exponentially. 
Although the axion can interact with the Standard Model photon, the photon acquires a thermal effective mass in the Universe filled with charged particles, which is much larger than the Hubble parameter, i.e. the axion mass at the onset of the oscillation. 
In that case, the gauge field production is kinematically prohibited.
Therefore, we assume that the gauge fields are hidden photons which are not thermalized at the onset of the axion oscillation.

To see the exponential growth, let us decompose the Fourier mode of the gauge fields into two circular polarization modes as ${\bf A}({\bf k},t) = A_+({\bf k},t) {\bf e}^+({\bf k})+A_-({\bf k},t) {\bf e}^-({\bf k})$ with the circular polarization bases, which satisfy $\hat{\bf k} \cdot {\bf e}^\pm = 0$ and $i\hat{\bf k} \times {\bf e}^\pm = \pm {\bf e}^\pm$ with $\hat{\bf k} \equiv {\bf k}/|{\bf k}|$.
In the linear approximation, the axion is assumed to be homogeneous in the equation of motion for the gauge fields. In this case, the dynamics of each circular polarization mode is determined by the following equation of motion,
\begin{eqnarray} \label{eq:gauge_linear}
 \ddot{A}_{\pm}  +H \dot{A}_{\pm} +\left(\frac{k^2}{a^{2}} \mp  \frac{k}{a} \frac{\alpha \dot\phi}{f} \right)  A_{\pm}=0.
\end{eqnarray}
This equation implies that, depending on the sign of $\dot\phi$, one of the two circular polarization modes can be tachyonic, leading to the exponential amplification of the gauge field amplitude.

\paragraph{Nonlinear dynamics}
-- Once the tachyonic instability turns on, 
the gauge fields grow exponentially and the energy density of the gauge fields eventually becomes comparable to that of the axion.
Then the gauge field production is saturated and the linear approximation is broken down. In particular, the produced gauge fields start to affect the axion dynamics through the right hand side in Eq.~(\ref{eq:axion_eom}), producing inhomogeneous modes of the axion. 
After that, the two polarization modes no longer evolve independently, but they are mixed through the interaction with nonzero mode axions.
In addition, the subsequent rescattering process significantly modifies the momentum distributions of both the axion and the gauge fields. Therefore, one needs to solve the nonlinear dynamics  
of the system to accurately compute the GW sources and the relic axion abundance.

\begin{figure}[tp]
\centering
\includegraphics [width = 8cm, clip]{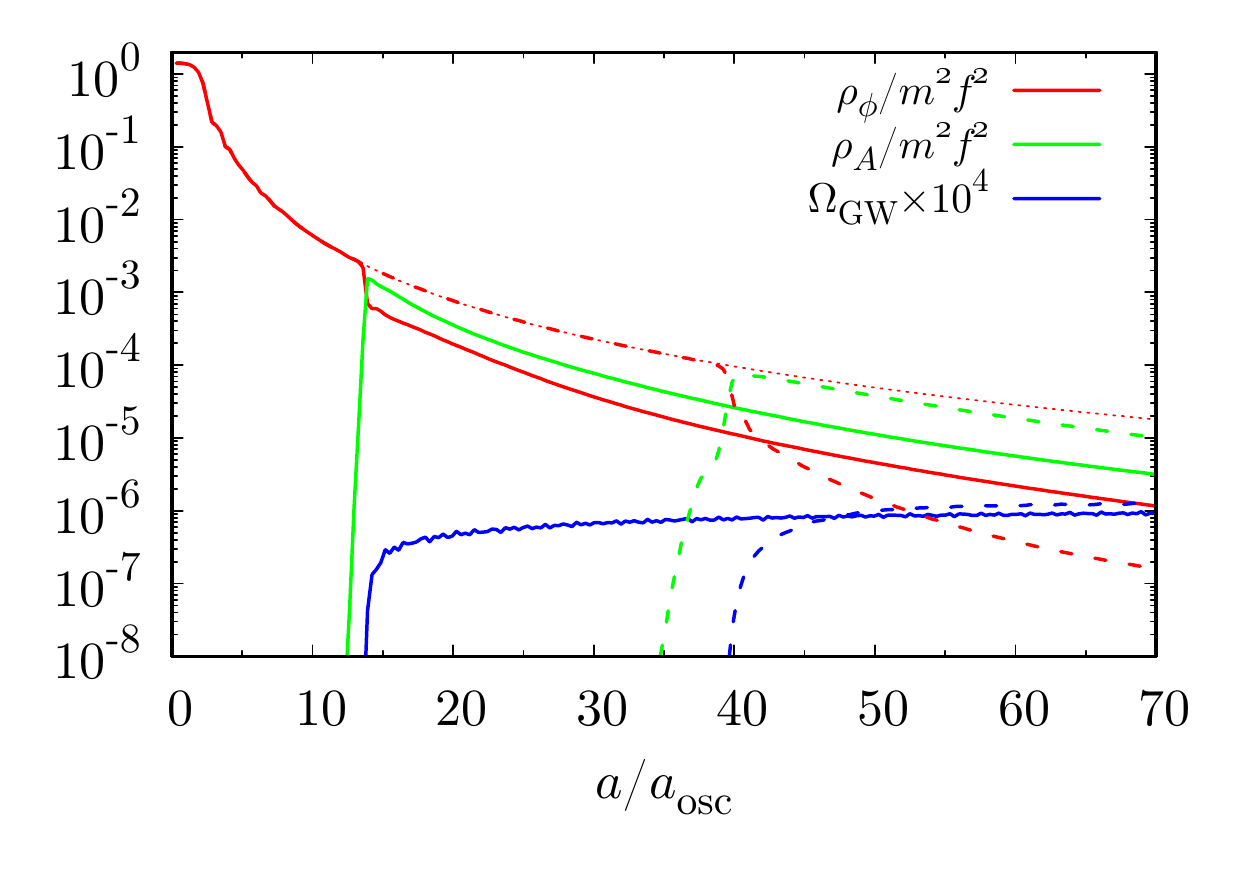}
\caption{
The evolution of the energy density of the axion (red), the gauge fields (green) and $\Omega_{\rm GW}\times 10^4$ (blue). For a comparison, we also show the adiabatic evolution without gauge field production by the thin dotted red line. We have taken $f=10^{16}$ GeV, $m=2\times 10^{-14}$ eV, $\theta_i= 2$ and $\alpha=30$ (solid) and 18 (dashed).
}
\label{fig:evolve}
\end{figure}

We have directly solved the equations of motion (\ref{eq:axion_eom}) and (\ref{eq:gauge_eom}) by performing numerical lattice simulations which enable us to analyze accurately the fully nonlinear dynamics. 
The number of grid points in the discretized lattice space is $256^3$ and the comoving box size is $(\pi/2)m^{-1}$ for $\alpha=18$ or $(\pi/4)m^{-1}$ for $\alpha=20,\,25,\,30$.
Fig.~\ref{fig:evolve} shows the evolution of the energy density of the axion, $\rho_\phi$ (red lines) and the gauge fields, $\rho_A$ (green lines) in terms of the scale factor divided by the one at the onset of the axion oscillation, defined by $H=m$. The figure shows that the growth of the gauge fields terminates when its energy density catches up that of the axion, namely $a/a_{\rm osc} \simeq 15$ for $\alpha=30$, which implies that the dynamics enters the nonlinear regime after that.

%%%%%%%%%%%%%%% FIGURE  %%%%%%%%%%%%%%%
%%%%%%%%%%%%%%%%%%%%%%%%%%%%%%%%%%%

\begin{figure}[tp]
\centering
\includegraphics [width = 8cm, clip]{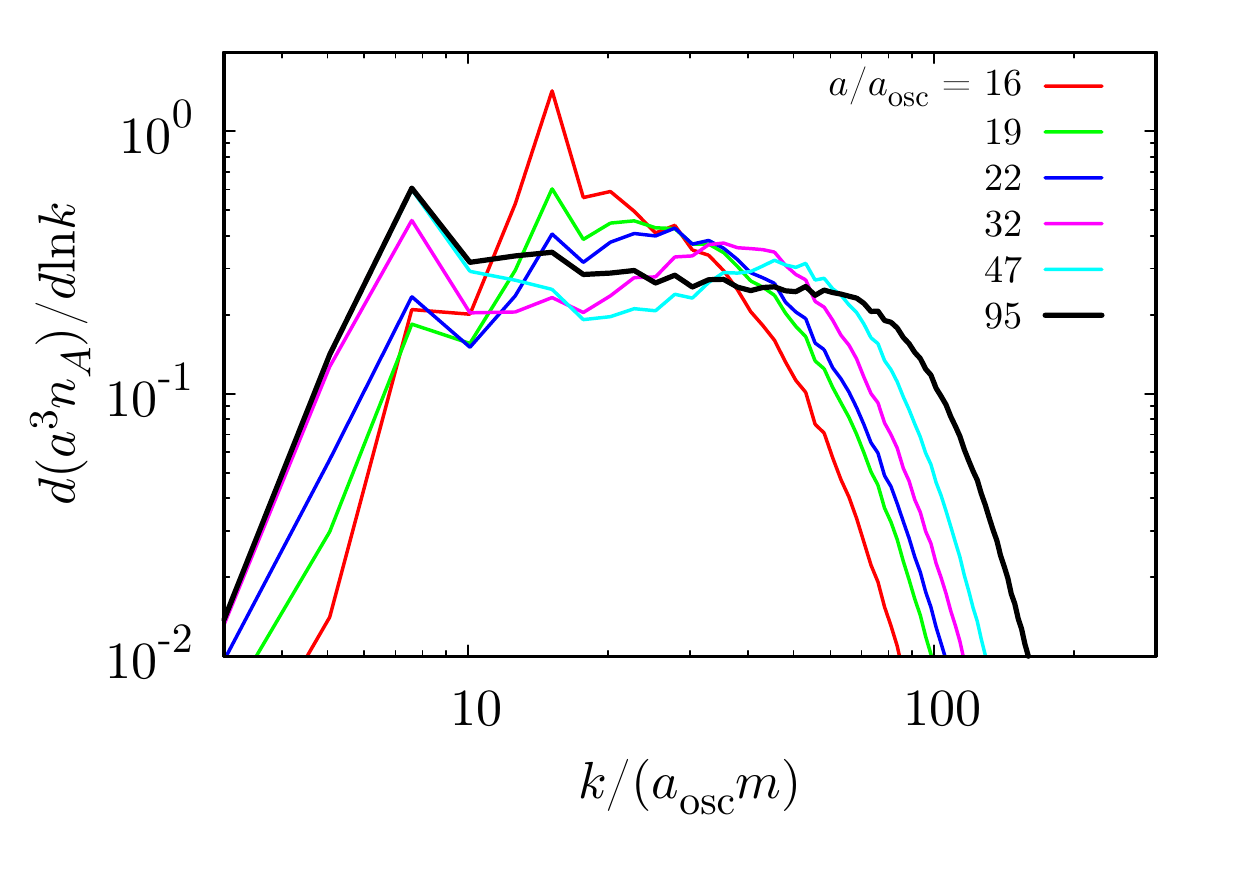}
\caption{
The evolution of the spectrum of comoving number density of the gauge fields ($a^3n_A)$ after the saturation normalized by $a_{\rm osc}^3 mf^2$. We have taken  $f=10^{16}$ GeV, $m=2\times 10^{-14}$ eV, $\theta_i= 2$, and $\alpha=30$.
}
\label{fig:dndlnk_alpha30}
\end{figure}

Fig.~\ref{fig:dndlnk_alpha30} shows the evolution of the comoving number density spectrum of the gauge fields after the gauge field production is saturated. The spectrum has a sharp peak soon after the saturation, i.e. at $a/a_{\rm osc} = 16$, but it is broaden and flatten as time goes on. It is nothing but the consequence of the nonlinear dynamics. As a result, the spectrum is significantly deformed from the one at the saturation, implying the significant departure from the one obtained by the linear analysis.

\begin{figure}[tp]
\centering
\includegraphics [width = 8cm, clip]{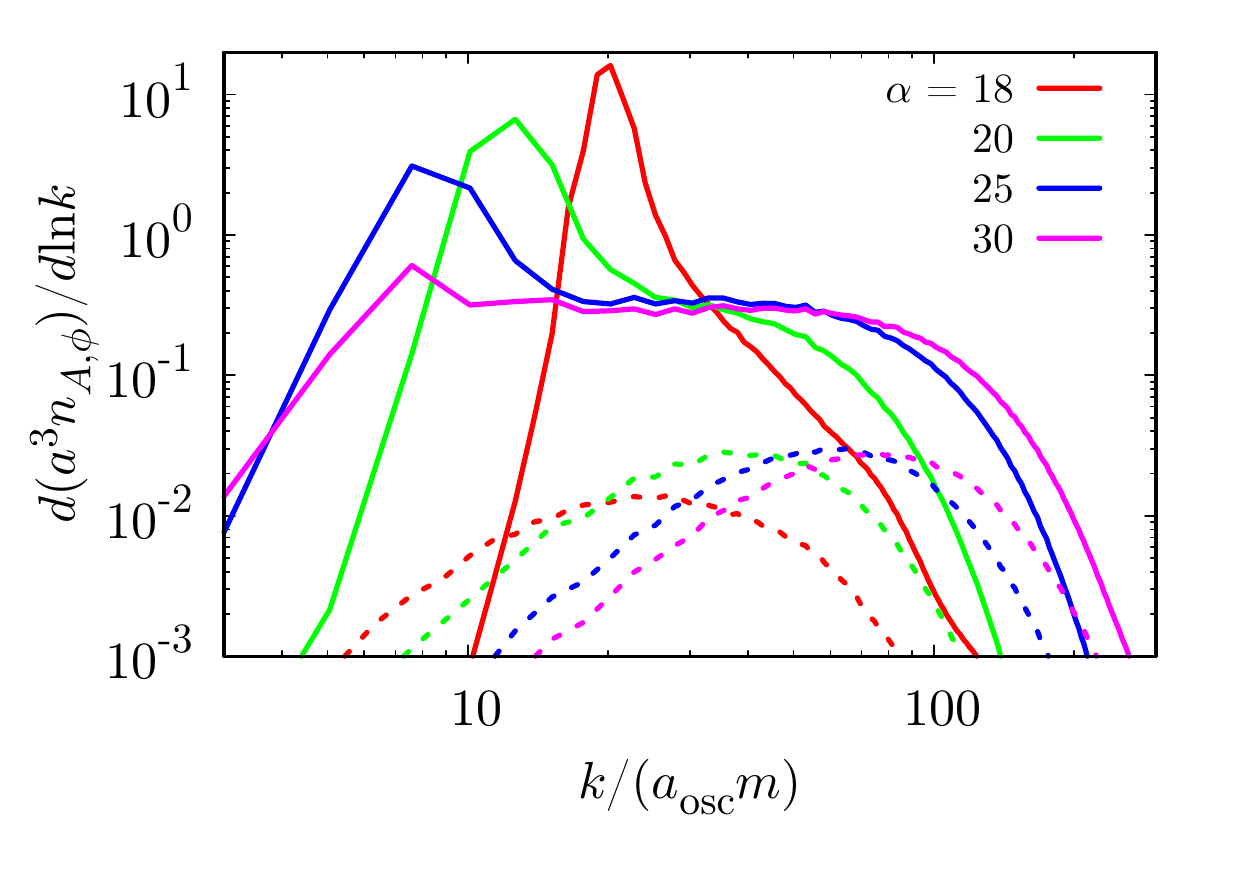}
\caption{
The $\alpha$-dependence of the resultant spectra of the comving number density of the gauge fields ($a^3n_A$, solid) and the axion ($a^3n_\phi$, dashed) normalized by $a_{\rm osc}^3 mf^2$. We have taken  $f=10^{16}$ GeV, $m=2\times 10^{-14}$ eV, $\theta_i= 2$, $\alpha=18$ (red), 20 (green), 25 (blue), and 30 (magenta).
}
\label{fig:dndlnk}
\end{figure}

Fig.~\ref{fig:dndlnk} shows the $\alpha$-dependence of the resultant comoving number density spectra of the gauge fields (solid lines) and the axion (dashed lines).
For $\alpha=18$, the sharp peak is kept in the spectrum, indicating that the nonlinear effect is not so efficient. 
On the other hand, the spectrum was significantly deformed through the nonlinear effect for $\alpha \geq 20$. 
Meanwhile, we have found that for $\alpha \leq 17$, the growth of the gauge fields stops before the saturation due to the cosmic expansion. 
The threshold value changes for different choices of $f$, $m$ and $\theta_i$.

%%%%%%%%%%%%%%% FIGURE  %%%%%%%%%%%%%%%
%%%%%%%%%%%%%%%%%%%%%%%%%%%%%%%%%%%

%%%%%%%%%%%%%%% FIGURE  %%%%%%%%%%%%%%%
%%%%%%%%%%%%%%%%%%%%%%%%%%%%%%%%%%%

\paragraph{Relic axion dark matter abundance}
-- 
Fig.~\ref{fig:evolve} also shows that the axion energy density drops suddenly at the satuation, leading to the suppression of the relic axion abundance compared to the case without gauge field productions (thin dotted red line). The suppression factor, $\epsilon$, can be as small as $\sim 10^{-2}$, as also pointed out in \cite{Kitajima:2017peg} in the QCD axion case.
Thus, we obtain the relic axion abundance at present as 
\begin{equation} \label{Abundance}
\Omega_a h^2 \sim 2.8 \epsilon \left(\frac{g_{*{\rm osc}}}{10} \right)^{-1/4} \left(\frac{m}{10^{-14}{\rm eV}} \right)^{1/2} \left(\frac{f \theta_i}{10^{16}{\rm GeV}} \right)^2,
\end{equation}
where, $g_{*{\rm osc}}$ is the effective relativistic degrees of freedom at the onset of the axion oscillation. Note that nonzero mode axions dominantly contribute to the abundance \cite{Kitajima:2017peg}.
If a string axion with $f \sim 10^{16}$ GeV accounts for the present dark matter component, 
the observed dark matter abundance determines the corresponding axion mass as $m\sim 10^{-14}$ eV for $\epsilon \sim 10^{-2}$.

\paragraph{Gravitational wave emission}
-- The explosively generated gauge fields can source the stochastic background of GWs,
which is given by the tensor component of the metric perturbation, $h_{ij}$, defined in the background flat-FRW Universe as follows
\begin{equation}
ds^2 = -dt^2 + a^2(\delta_{ij}+h_{ij}) dx^i dx^j.
\end{equation}
The perturbed Einstein equation for the tensor mode determines the evolution of the GW,
\begin{equation}
\ddot{h}_{ij}+3H\dot{h}_{ij}-\frac{1}{a^2} \nabla^2 h_{ij} = \frac{2}{M_P^2} \Pi^{\rm TT}_{ij},
\end{equation}
where $\Pi^{\rm TT}_{ij}$ is the transverse-traceless part of the anisotropic stress.
In the system of our interest, the anisotropic stress is given by the traceless part of 
\begin{equation}
\Pi_{ij} = -\frac{1}{a^2}\partial_i \phi \partial_j \phi + \frac{1}{a^2}E_i E_j + \frac{1}{a^2} B_i B_j.
\end{equation}
where $E_i = \dot{A}_i$ and $B_i = \epsilon_{ijk} \partial_j A_k/a$ are respectively the electric and magnetic components of the gauge fields.

\begin{figure}[tp]
\centering
\includegraphics [width = 8cm, clip]{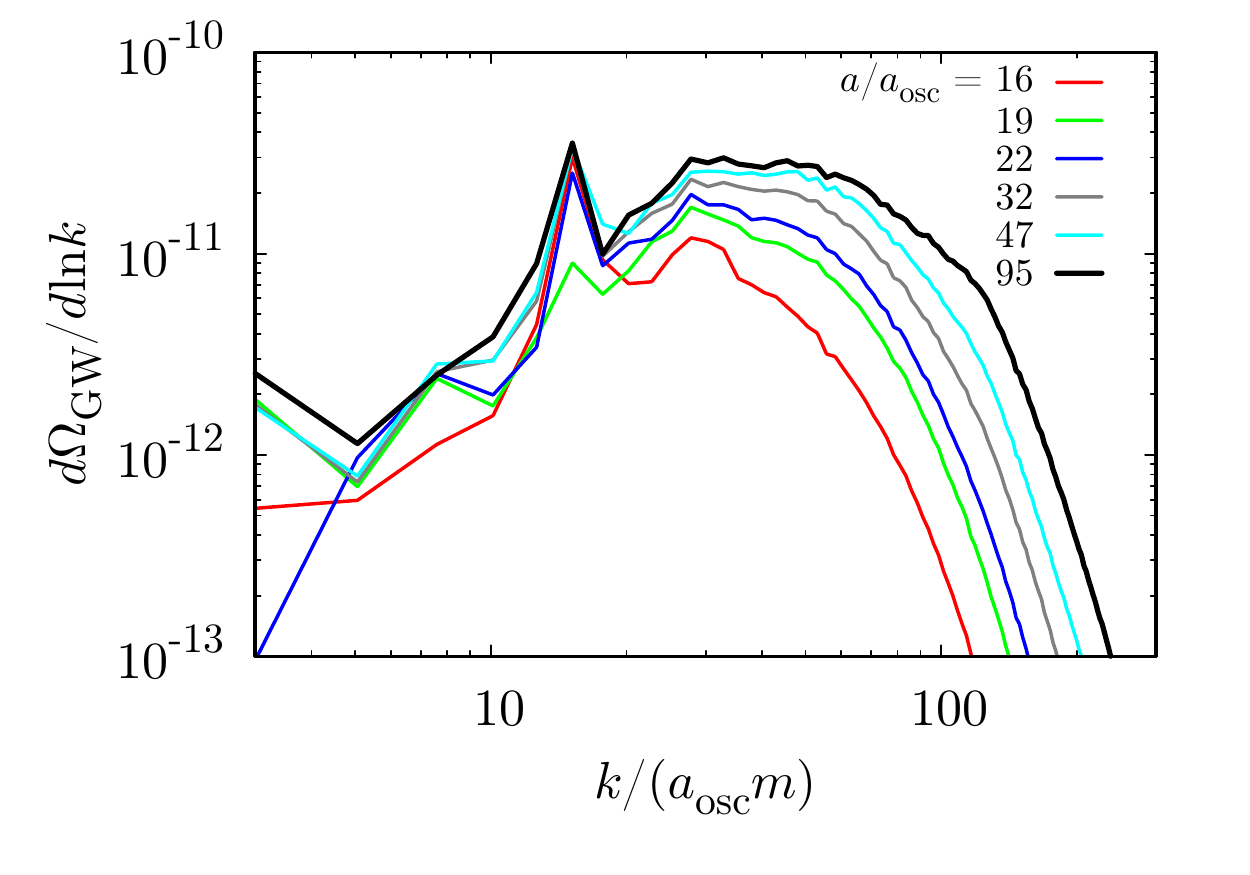}
\caption{
The evolution of the spectrum of $\Omega_{\rm GW}$ after the saturation for the same setup as in Fig.~\ref{fig:dndlnk_alpha30}.
}
\label{fig:OmegaGW_alpha30}
\end{figure}

\begin{figure}[tp]
\centering
\includegraphics [width = 8cm, clip]{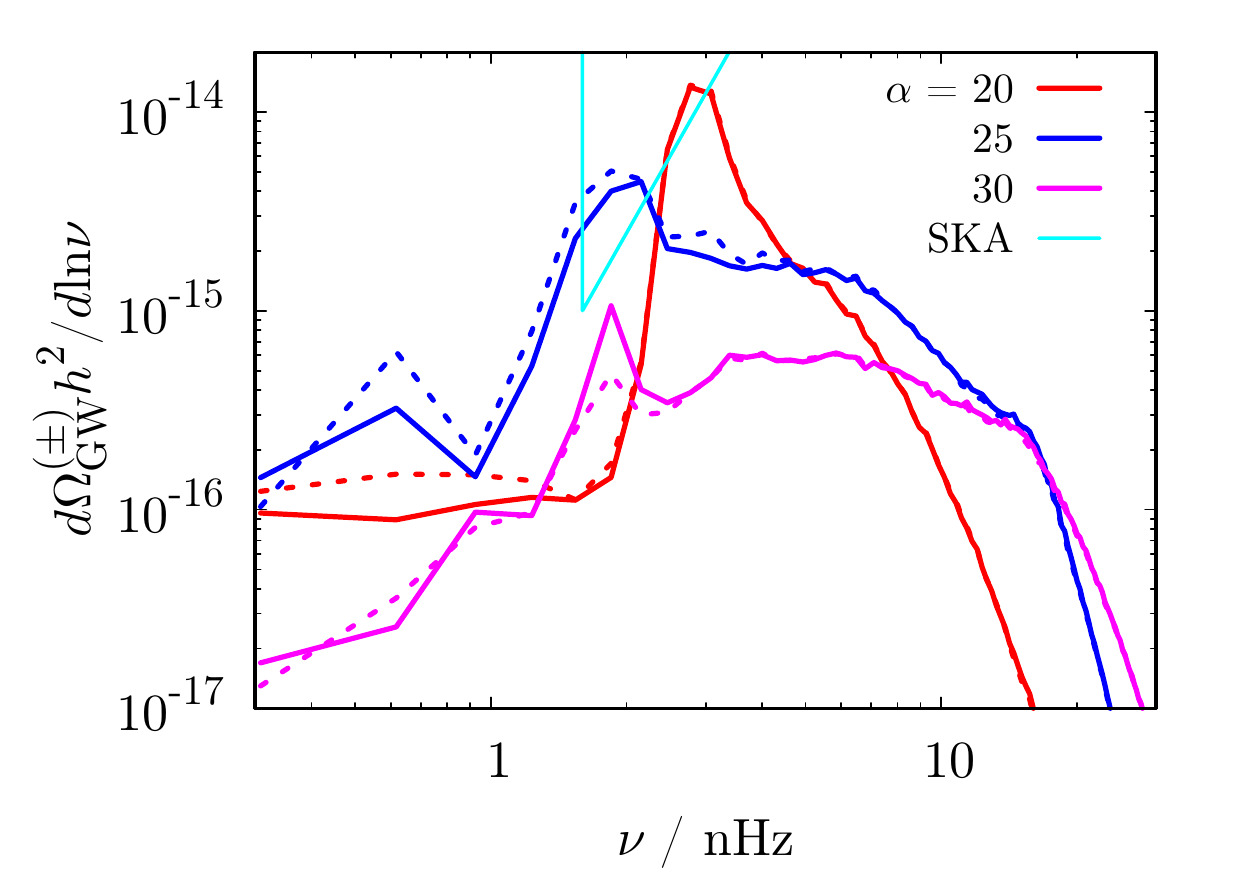}
\caption{
The $\alpha$ dependence of the resultant spectrum of $\Omega_{\rm GW}h^2$ decomposed into two circular polarization modes, $\Omega_{\rm GW}^{(+)}h^2$ (solid line) and $\Omega_{\rm GW}^{(-)}h^2$ (dashed line). We have taken $f=10^{16}$ GeV, $\theta_i=2$, $m=2\times 10^{-14}$ eV and $\alpha=20$ (red), 25 (blue), 30 (magenta).
}
\label{fig:OmegaGW}
\end{figure}

First, let us make a crude estimation of the peak amplitude 
of the GWs sourced by gauge fields or inhomogeneous axions which are amplified by the oscillating axion. 
Assuming for simplicity that the dominant fraction of the GWs is emitted around the saturation, the peak amplitude of the GW is roughly estimated as
\begin{equation} \label{eq:estimate_h}
\frac{k^2}{a_{\rm em}^2} h_{ij}(t_{\rm em}) \sim \frac{\rho_{\rm src}(t_{\rm em})}{M_P^2} \sim \left( \frac{mf\theta_i}{M_P} \right)^2 \left(\frac{a_{\rm osc}}{a_{\rm em}} \right)^3,
\end{equation}
where $a_{\rm em}$ is the scale factor at the emission, $t=t_{\rm em}$, and $\rho_{\rm src}$ is the energy density of the source fields, which becomes comparable to the energy density of the homogeneous mode of the axion at $t_{\rm em}$. Then, one obtains the density parameter of the GW, $\Omega_{\rm GW} = M_P^2 \langle \dot{h}_{ij} \dot{h}_{ij} \rangle/(4\rho_{\rm cr})$ (with the critical density $\rho_{\rm cr}$), at the emission as \cite{Kitajima:2018zco}
\begin{equation}
\Omega_{\rm GW}(t_{\rm em}) \sim \left(\frac{k}{a_{\rm em} m}\right)^{-2} \left(\frac{f\theta_i}{M_P}\right)^4 \left(\frac{a_{\rm osc}}{a_{\rm em}} \right)^2, \label{Exp:OmegaGW_estimation}
\end{equation}
and at present as $\Omega_{\rm GW}h^2 = \kappa \Omega_r h^2 \Omega_{\rm GW}(t_{\rm em})$, where $\Omega_r h^2 \simeq 4.15\times 10^{-5}$ is the present density parameter of radiation component and $\kappa$ is given by $\kappa \simeq 1.8 g_{*{\rm em}}/g_{* {\rm S\,em}}^{4/3}$ with $g_{*{\rm em}}$ and $g_{* {\rm S\,em}}$ being the effective degrees of freedom in energy density and entropy density at the emission.\footnote{
Eqs. (\ref{eq:estimate_h}) and (\ref{Exp:OmegaGW_estimation}) are more generically applicable to the GW production during radiation domination whose source fields are generated by homogeneously oscillating axions with general $m$ and $f$ as addressed in the GW forest scenario ~\cite{Kitajima:2018zco}.} 
Note that the typical wavenumber of emitted GWs is $k/a_{\rm em} \sim m$. For example, for $f=10^{16}$ GeV, $\theta_i=2$, $m=2 \times 10^{-14}$ eV and $\alpha = 30$, we found $a_{\rm em}/a_{\rm osc} \simeq 15$ from Fig.~\ref{fig:evolve} and then we obtain $\Omega_{\rm GW}(t_{\rm em}) \sim 2 \times 10^{-11}$ ($\Omega_{\rm GW}h^2 \sim 7 \times 10^{-16}$) which shows an acceptable agreement with the numerical result (see Fig.~\ref{fig:OmegaGW_alpha30} and \ref{fig:OmegaGW}).

The GW frequency at present is related to the axion mass as follows \cite{Soda:2017dsu, Kitajima:2018zco}
\begin{equation}
\nu = \frac{k}{2\pi a_0} \sim 0.1 {\rm nHz} \frac{g_{*{\rm osc}}^{1/4}}{g_{*S{\rm osc}}^{1/3}} \frac{k}{a_{\rm osc} m} \left( \frac{m}{10^{-14} {\rm eV} } \right)^{\frac{1}{2}}  \,, \label{Exp:frequency_estimation}
\end{equation} 
with $a_0$ the present scale factor and $g_{*S{\rm osc}}$ the effective relativistic degrees of freedom in entropy density at the onset of the axion oscillation. 
For  $f \sim 10^{16}$ GeV, the axion with $m \sim 10^{-14}$ eV, which gives the observed dark matter abundance, predicts the GWs in nHz range with $\Omega_{\rm GW} h^2  \sim 10^{-15}$ for $a_{\rm osc}/a_{\rm em} = O(0.1)$.

To obtain the accurate GW spectrum in this scenario, we have numerically solved the evolution of GWs together with the dynamics of the axion and the gauge fields by the lattice simulation formulated in \cite{GarciaBellido:2007af}.
The evolution of $\Omega_{\rm GW}$ is shown by the blue lines in Fig.~\ref{fig:evolve}. The generation of the GWs 
continues even after the saturation, being affected by the nonlinear dynamics of the gauge fields and the axion such as the rescattering. 
In particular, as shown in Fig.~\ref{fig:OmegaGW_alpha30}, which shows the evolution of the spectrum of $\Omega_{\rm GW}$, i.e. $d\Omega_{\rm GW}/d\ln k$, for $\alpha=30$, the spectrum is broaden and the maximum amplitude becomes larger after the system enters the nonlinear regime. The former is because the spectrum of the gauge fields becomes broader as shown in Fig.~\ref{fig:dndlnk_alpha30}.
In most cases, compared to the numerical result, (\ref{Exp:OmegaGW_estimation}) underestimates the amount of GWs because the emission after the saturation is ignored.

Fig.~\ref{fig:OmegaGW} shows the prediction of the present GW spectrum together with the sensitivity of SKA (the cyan line)~\cite{Janssen:2014dka} in the case where the axion gives a consistent value with the observed dark matter abundance.
Here, $\Omega_{\rm GW}^{(\pm)}$ denotes the contribution of either of two circular polarization modes to the total $\Omega_{\rm GW}$. Fig.~\ref{fig:OmegaGW}  indicates that the GWs from the axion dark matter are detectable by SKA, providing a new window for axion dark matter search.  The spectrum has a sharp peak for $\alpha=20$, and as $\alpha$ increases, the width of the spectrum becomes broader. Accordingly, the height of the spectrum drops by ${\cal O}(0.1)$. Therefore this model can predict both a highly peaked spectrum and a broadly extended spectrum depending on $\alpha$.

Let us address whether the reported signal by NANOGrav \cite{Arzoumanian:2020vkk} can be accounted for in this scenario.
It requires $d\Omega_{\rm GW}h^2/d\ln\nu \sim 10^{-9}$ in nHz range.
Since the density of emitted GWs roughly scales as $\Omega_{\rm GW} \propto f^4$ and $\Omega_{\rm GW} \sim 10^{-15}$\,-\,$10^{-14}$ for $f = 10^{16}$ GeV (see Fig.~\ref{fig:OmegaGW}), one needs $f \sim 2$\,-\,$4\times 10^{17}$ GeV to explain the NANOGrav data.
In that case, however, the relic axion abundance becomes two or three orders of magnitude larger than the observed dark matter abundance.
Then, one needs further suppression mechanism for this scenario to work with such a large decay constant.

%%%%%%%%%%%%%%% FIGURE  %%%%%%%%%%%%%%%
%%%%%%%%%%%%%%%%%%%%%%%%%%%%%%%%%%%

%%%%%%%%%%%%%%% FIGURE  %%%%%%%%%%%%%%%
%%%%%%%%%%%%%%%%%%%%%%%%%%%%%%%%%%%

\begin{figure}[tp]
\centering
\includegraphics [width = 8cm, clip]{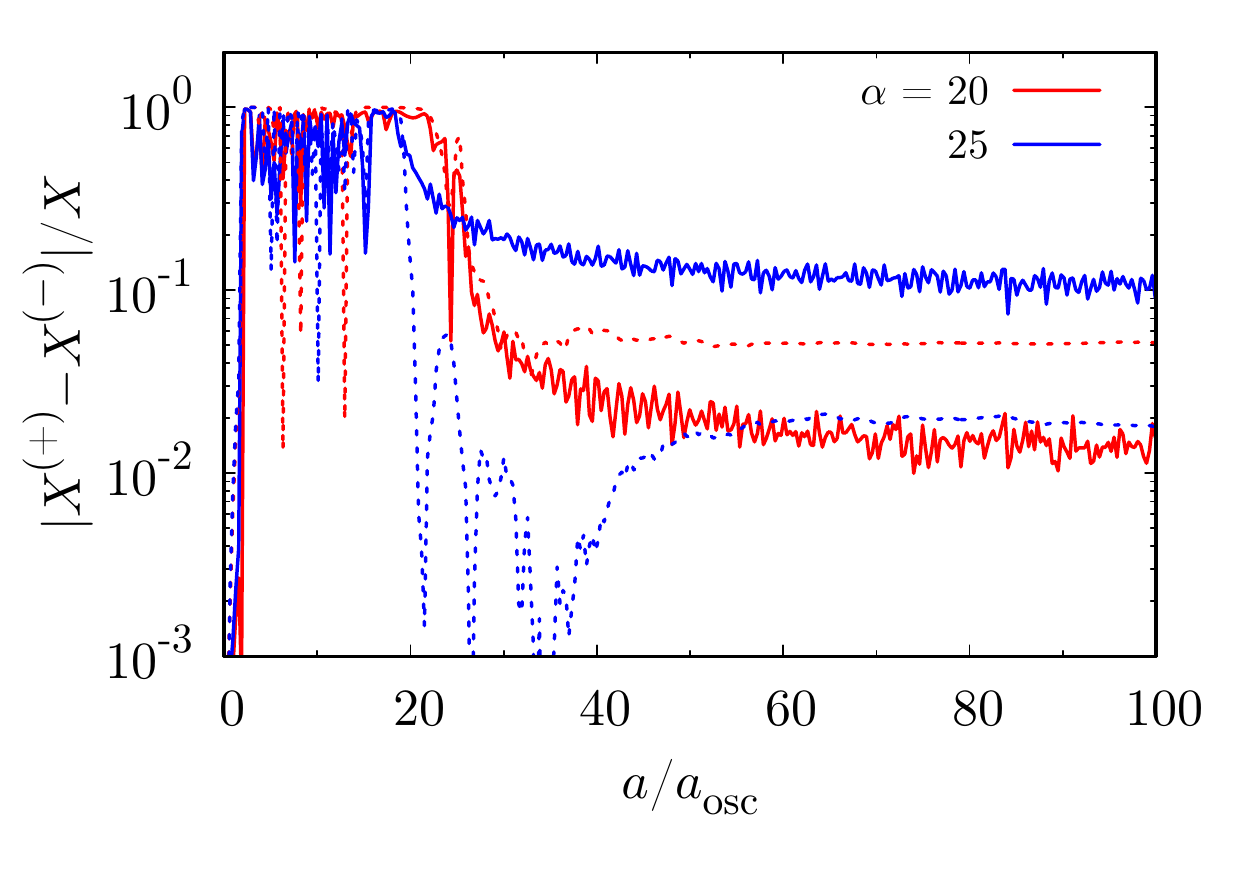}
\caption{
The evolution of the circular polarization asymmetry for GWs ($X=\Omega_{\rm GW}$, solid lines) and gauge fields ($X=\rho_{A}$, dashed lines). We have taken $f=10^{16}$ GeV, $\theta_i=2$, $m=2\times 10^{-14}$eV, and $\alpha=20$ (red) and $25$ (blue).
}
\label{fig:evolve_pol}
\end{figure}

As is shown in Fig.~\ref{fig:OmegaGW}, this scenario predicts not only the detectable GW signal but also the circular polarization of the GW.
Fig.~\ref{fig:evolve_pol} shows the time evolution of the difference between two circular polarization modes of the density parameter of the GW, $\Omega_{\rm GW}$ and the energy density of the gauge fields, $\rho_{A}$. 
The predicted asymmetry of the circular polarization is 1\,-\,10 \% depending on the coupling.
It is possible in principle to detect the circular polarization with pulsar timing arrays by observing an anisotropy of GWs~\cite{Kato:2015bye,Belgacem:2020nda}.

%%%%%%%%%%%%%%% FIGURE  %%%%%%%%%%%%%%%
%%%%%%%%%%%%%%%%%%%%%%%%%%%%%%%%%%%

\paragraph{Discussion}
--
In this letter, we focus only on the axion playing a role of the dark matter. Specifically, for the GUT scale decay constant, i.e. $f \sim 10^{16}$ GeV, the axion mass should be $m \sim 10^{-14}$ eV corresponding to nHz range of the emitted GWs.
However, axions with different masses might exist in the context of the string axiverse. In that case, GWs in multi-frequency bands can also be produced in the cosmological history, dubbed GW forest~\cite{Soda:2017dsu, Kitajima:2018zco} (see also Refs.~\cite{Machado:2018nqk, Arvanitaki:2019rax}).
Although axions in some mass ranges have to decay after the GW emission to avoid overproduction, the GW emission process for other frequency bands follows more or less the same story as the one in nHz band studied in this letter, predicting the circularly polarized GW forest for a certain range of $\alpha$.

When the potential is different from the cosine form, e.g., having a plateau region, the nature of the gauge field production and accordingly the spectra can be significantly modified. We leave such extensions for a future study.

\acknowledgements
We thank Wolfram Ratzinger and Pedro Schwaller for helpful comments.
N.~K. is supported by JSPS KAKENHI Grant Number JP18H01243, JP19K14708 and 20H01894. N.~K. and Y.~U. are supported by JSPS KAKENHI Grant Number JP19H01894. J.~S. was in part supported by JSPS KAKENHI
Grant Numbers JP17H02894 and JP20H01902. Y.~U. is also supported by the Deutsche Forschungsgemeinschaft (DFG, German Research Foundation) - Project number 315477589 - TRR 211. The authors thank Yukawa Institute for Theoretical Physics at Kyoto University. Discussions during the YITP workshop YITP-T-19-02 on "Resonant instabilities in cosmology" were useful to complete this work.

\bibliography{refst}

\end{document}